\documentstyle[thmsa,a4,12pt,sw20lart]{article}

\input tcilatex
\begin{document}

\title{{\bf Controlled Quantum Teleportaion}$^{\thanks{%
The project supported by National Natural Science Foundation of China.}}$}
\author{Jindong Zhou$^2$, Guang Hou$^2$, Shenjun Wu$^2$ and Yongde Zhang$^{1,2}$ \\
$^1${\em CCAST (World Laboratory), P.O. Box 8730,}\\
{\em Beijing 100080, P.R. China.}\\
$^2${\em Department of Modern Physics,}\\
{\em University of Science and Technology of China.}\\
{\em Hefei, 230027, P.R. China.}}
\date{
\today%
}
\maketitle

\begin{abstract}
A theoretical scheme for controlled quantum teleportation is presented,
using the entanglement property of {\bf GHZ} state.
\end{abstract}

\section{Introduction}

In the field of quantum computation and communication, quantum channel plays
an important role. In 1993, {\em Bennet} et.al.$^{\cite{sixmen}}$ presented
the theory of quantum teleportation, which provides a theoretical basis for
the construction of quantum channels. The first experiment of quantum
teleportation was accomplished by {\em Boumeester} et.al.$^{\cite{teleport}}$
in 1997, which was generally acknowledged as a milestone in the field of
quantum information. In that work, the technique of {\em Bell}-state analysis%
$^{\cite{bellanaly1,bellanaly2}}$ was utilized. In 1999, the first
three-photon entangle state({\bf GHZ} state$^{\cite{ghztheory}}$) was
experimentally realized$^{\cite{ghzexpe1,ghzexpe2}}$.

In this paper, the entanglement property of {\bf GHZ} state is utilized to
design a theoretical scheme for controlled quantum teleportation. According
to the scheme, a  third side is included, so that the quantum channel is
supervised by this additional side. The signal state cannot be transmitted
unless all three sides agree to cooperate.

\section{Controlled quantum teleportation}

Suppose {\em Alice}, {\em Bob} and {\em Charlie} share a {\bf GHZ} state$^{%
\cite{ghztheory}}$,
\[
\left| \psi \right\rangle _{ABC}=\frac 1{\sqrt{2}}\left( \left|
000\right\rangle +\left| 111\right\rangle \right) _{ABC}
\]
and {\em Alice} is to teleport an unknown signal state to {\em Bob}. The
signal was originally carried by qubit $D$,
\[
\left| \psi \right\rangle _D=\alpha \left| 0\right\rangle _D+\beta \left|
1\right\rangle _D
\]

The state for the whole system(four qubits) can be expressed as:
\begin{eqnarray*}
\left| \psi \right\rangle _{DABC} &=&\left( \alpha \left| 0\right\rangle
+\beta \left| 1\right\rangle \right) _D\otimes \frac 1{\sqrt{2}}\left(
\left| 000\right\rangle +\left| 111\right\rangle \right) _{ABC} \\
&=&\frac 12\cdot \frac 1{\sqrt{2}}\left( \left| 00\right\rangle +\left|
11\right\rangle \right) _{DA}\otimes \left( \alpha \left| 00\right\rangle
+\beta \left| 11\right\rangle \right) _{BC} \\
&&+\frac 12\cdot \frac 1{\sqrt{2}}\left( \left| 00\right\rangle -\left|
11\right\rangle \right) _{DA}\otimes \left( \alpha \left| 00\right\rangle
-\beta \left| 11\right\rangle \right) _{BC} \\
&&+\frac 12\cdot \frac 1{\sqrt{2}}\left( \left| 01\right\rangle +\left|
10\right\rangle \right) _{DA}\otimes \left( \beta \left| 00\right\rangle
+\alpha \left| 11\right\rangle \right) _{BC} \\
&&+\frac 12\cdot \frac 1{\sqrt{2}}\left( \left| 01\right\rangle -\left|
10\right\rangle \right) _{DA}\otimes \left( -\beta \left| 00\right\rangle
+\alpha \left| 11\right\rangle \right) _{BC}
\end{eqnarray*}

Now {\em Alice} perform a {\em Bell}-state measurement$^{\cite
{bellanaly1,bellanaly2}}$ on qubits $DA$. After that, she will broadcast the
result of her measurement, so that qubits $BC$ can be transformed(According
to the four possible results $\frac 1{\sqrt{2}}\left( \left| 00\right\rangle
+\left| 11\right\rangle \right) _{DA}$, $\frac 1{\sqrt{2}}\left( \left|
00\right\rangle -\left| 11\right\rangle \right) _{DA}$, $\frac 1{\sqrt{2}%
}\left( \left| 01\right\rangle +\left| 10\right\rangle \right) _{DA}$ and $%
\frac 1{\sqrt{2}}\left( \left| 01\right\rangle -\left| 10\right\rangle
\right) _{DA}$, the corresponding transformations are $I_B\otimes I_C$, $%
I_B\otimes \left( \left| 0\right\rangle \left\langle 0\right| -\left|
1\right\rangle \left\langle 1\right| \right) _C$, $\left( \left|
0\right\rangle \left\langle 1\right| +\left| 1\right\rangle \left\langle
0\right| \right) _B\otimes \left( \left| 0\right\rangle \left\langle
1\right| +\left| 1\right\rangle \left\langle 0\right| \right) _C$ and $%
\left( \left| 0\right\rangle \left\langle 1\right| +\left| 1\right\rangle
\left\langle 0\right| \right) _B\otimes \left( \left| 0\right\rangle
\left\langle 1\right| -\left| 1\right\rangle \left\langle 0\right| \right) _C
$, respectively.) by {\em Bob} and/or {\em Charlie} to a common form:
\[
\left| \psi \right\rangle _{BC}=\left( \alpha \left| 00\right\rangle +\beta
\left| 11\right\rangle \right) _{BC}
\]

Considering this state, we can see that, at this moment, neither  {\em Bob}
nor {\em Charlie} can obtain the original signal state $\alpha \left|
0\right\rangle +\beta \left| 1\right\rangle $ without the cooperation of the
other one.

If {\em Charlie} would like to help {\em Bob} for the teleportation, he
should just measure his portion of $BC$, namely qubit $C$, on the bases of $%
\frac 1{\sqrt{2}}\left( \left| 0\right\rangle +\left| 1\right\rangle \right)
_C$ and $\frac 1{\sqrt{2}}\left( \left| 0\right\rangle -\left|
1\right\rangle \right) _C$, and transfer the result of his measurement to 
{\em Bob} via a classical channel. Here the state of qubits $BC$ can be
written as:
\begin{eqnarray*}
\left| \psi \right\rangle _{BC} &=&\left( \alpha \left| 00\right\rangle
+\beta \left| 11\right\rangle \right) _{BC} \\
&=&\frac 1{\sqrt{2}}\cdot \left( \alpha \left| 0\right\rangle +\beta \left|
1\right\rangle \right) _B\otimes \frac 1{\sqrt{2}}\left( \left|
0\right\rangle +\left| 1\right\rangle \right) _C \\
&&+\frac 1{\sqrt{2}}\cdot \left( \alpha \left| 0\right\rangle -\beta \left|
1\right\rangle \right) _B\otimes \frac 1{\sqrt{2}}\left( \left|
0\right\rangle -\left| 1\right\rangle \right) _C
\end{eqnarray*}

As soon as {\em Bob} is informed {\em Charlie}'s result, he can perform an
appropriate unitary transformation(According to the two possible results $%
\frac 1{\sqrt{2}}\left( \left| 0\right\rangle +\left| 1\right\rangle \right)
_C$ and $\frac 1{\sqrt{2}}\left( \left| 0\right\rangle -\left|
1\right\rangle \right) _C$, the corresponding transformations are $I_B$ and $%
\left( \left| 0\right\rangle \left\langle 0\right| -\left| 1\right\rangle
\left\langle 1\right| \right) _B$, respectively.) on qubit $B$ to obtain the
original signal state,
\[
\left| \psi \right\rangle _B=\alpha \left| 0\right\rangle _B+\beta \left|
1\right\rangle _B
\]

The feature of this scheme is that teleportation between two sides depends
on the agreement of the third side. It is therefore named ''Controlled
Quantum Teleportation''.

\section{Conclusion}

The difference of the scheme presented in this paper with the original
quantum teleportation scheme$^{\cite{sixmen}}$ is that a third side({\em %
Charlie}) is included, who may participate the process of quantum
teleportation as a supervisor. Without the cooperation(permission) of {\em %
Charlie}, {\em Bob} cannot get the signal state from {\em Alice} by himself.

This property of the scheme can be utilized to construct controlled quantum
channels, which may be useful in the future quantum computers.

\end{document}